\documentclass[pra,aps,a4paper,twocolumn,superscriptaddress,nofootinbib]{revtex4-2}

\usepackage[T1]{fontenc}
\usepackage[utf8]{inputenc}
\usepackage{amssymb}
\usepackage{amsmath}
\usepackage{graphicx}
\usepackage{bbm}
\usepackage{psfrag}
\usepackage{hyperref}
\usepackage{braket}
\usepackage[normalem]{ulem}
\usepackage[caption=false]{subfig} 
\usepackage{ragged2e} 
\DeclareCaptionJustification{justified}{\justifying}
\usepackage{float}

\usepackage{xcolor}
\hypersetup{
    colorlinks=true,
    citecolor=blue,
    linkcolor=blue,
    filecolor=magenta,
    urlcolor=blue}
\definecolor{ccqqqq}{rgb}{1,0,0}
\definecolor{uuuuuu}{rgb}{0.26666666666666666,0.26666666666666666,0.26666666666666666}
\definecolor{qqwwzz}{rgb}{0,0,1}
\usepackage{float}
\usepackage{tikz}
\usetikzlibrary{quantikz2}
\usepackage{quantikz}

\newcommand{\beq}{\begin{equation}}
\newcommand{\eeq}{\end{equation}}
\newcommand{\bea}{\begin{eqnarray}}
\newcommand{\eea}{\end{eqnarray}}
\newcommand{\bit}{\begin{itemize}}
\newcommand{\eit}{\end{itemize}}

\def\Tr{{\rm Tr}}

\usepackage{soul} 
\usepackage[normalem]{ulem}

\usepackage{orcidlink}

\newcommand{\e}{\epsilon}

\renewcommand{\t}{\tau}

\usepackage{geometry}
\newgeometry{includefoot,left=2cm,right=2cm,bottom=1cm,top=2cm}

\newcommand{\ie}{{\it i.e.,}\ }

\makeatletter
\DeclareRobustCommand{\cev}[1]{%
  \mathpalette\do@cev{#1}%
}
\newcommand{\do@cev}[2]{%
  \fix@cev{#1}{+}%
  \reflectbox{$\m@th#1\vec{\reflectbox{$\fix@cev{#1}{-}\m@th#1#2\fix@cev{#1}{+}$}}$}%
  \fix@cev{#1}{-}%
}
\newcommand{\fix@cev}[2]{%
  \ifx#1\displaystyle
    \mkern#23mu
  \else
    \ifx#1\textstyle
      \mkern#23mu
    \else
      \ifx#1\scriptstyle
        \mkern#22mu
      \else
        \mkern#22mu
      \fi
    \fi
  \fi
}
\makeatother

\begin{document}

\author{Tal Schwartzman}\email{tal.schwartzman@cfa.harvard.edu\\ \\}

\affiliation{ITAMP, Center for Astrophysics {\normalfont\textbar}  Harvard \& Smithsonian, Cambridge, Massachusetts 02138, USA}
\author{Torsten V. Zache}\email{torsten.zache@uibk.ac.at}
\affiliation{Institute for Quantum Optics and Quantum Information, Austrian Academy of Sciences, Innsbruck, 6020, Austria}
\affiliation{Institute for Theoretical Physics, University of Innsbruck, Innsbruck, 6020, Austria}

\author{Hannes Pichler}
\affiliation{Institute for Quantum Optics and Quantum Information, Austrian Academy of Sciences, Innsbruck, 6020, Austria}
\affiliation{Institute for Theoretical Physics, University of Innsbruck, Innsbruck, 6020, Austria}

\author{H. R. Sadeghpour}

\affiliation{ITAMP, Center for Astrophysics {\normalfont\textbar} Harvard \& Smithsonian, Cambridge, Massachusetts 02138, USA}

\title{Imaginary time evolution and ground state preparation \\ using unitary multi-copy protocols}

\begin{abstract} 
Efficient low-energy state preparation is a key objective in quantum computation and quantum simulation. Quantum imaginary-time evolution replaces real-time dynamics with imaginary-time dynamics, exponentially suppressing higher-energy eigenstates.
We introduce deterministic unitary protocols that approximate imaginary-time evolution for ground state preparation. The protocols require multiple copies of the system, real-time evolution under the system Hamiltonian, and controlled-SWAP operations (or more general SWAP-generated unitaries). Our analysis focuses on two concrete circuit families: a tree architecture with provable polynomial-in-depth convergence but rapidly growing width, and a compact “hedge” architecture that achieves comparable accuracy with only polynomial width in a heuristic construction supported by numerics. Numerical evidence indicates that mid-circuit post-selection can accelerate convergence with practical success probabilities. Separately, we demonstrate that circuit volume can be traded for the shot complexity of post-circuit observable estimation in the ground state preparation setting. Finally, we outline concrete platform-specific implementations in which multi-copy registers
and SWAP-mediated couplings are natural, illustrating how these hybrid analog-digital circuits
can complement existing state-preparation methods in the near term.
\end{abstract}

\maketitle

\color{black}

\section{Introduction}
Preparing ground states is one of the central goals of quantum computation and quantum simulation. 
From a physics perspective, ground states and low-lying excitations determine phases of quantum matter and its emergent low-temperature behavior, and they underpin electronic-structure calculations in chemistry and materials \cite{Sachdev_2011, mcardle2020quantum}. From an algorithmic perspective, computational tasks can be encoded into the ground state preparation of a suitable final Hamiltonian \cite{farhi2000quantum, aharonov2008adiabatic}. At the same time, in the worst case, preparing ground states of local Hamiltonians is intractable, as the local Hamiltonian problem is complete for the complexity class QMA \cite{kempe2004complexity}. However, there is a gap between worst-case hardness and the case of physically relevant models that makes the problem well motivated.

Many classical methods exploit the entanglement structure of physically relevant systems to construct efficient variational ansatz, most notably tensor networks and related schemes \cite{white1992density, schollwock2011density, orus2014practical}. Their performance, however, depends strongly on the entanglement structure and geometry, and it can degrade rapidly outside the area-law regime or in higher dimensions. {On the hardware stack, several paradigms for ground state preparation have been developed. Notable examples include approaches based on quantum phase estimation and quantum singular value transformation, as well as adiabatic state preparation, dissipative state engineering, and algorithmic cooling \cite{ge2019faster, gilyen2019quantum, albash2018adiabatic, lin2020near, dong2022ground, verstraete2009quantum,  QVC2019, roy2020measurement, AC2005, HBAC2015, AC2025, cubitt2023dissipative, mi2024stable, ding2024single, lloyd2025quasiparticle, ding2025end, zhan2026rapid}. In practice, one faces a compromise between theoretical control and experimental implementation. Provably accurate protocols typically demand substantial coherent depth and resources tied to spectral resolution or small gaps, whereas hardware-friendly cooling and dissipation can be simpler but are constrained by relaxation (mixing) times, and general high-fidelity pure-state guarantees often require additional structure or assumptions. It is therefore valuable to develop additional ground state preparation protocols compatible with near-term hardware while achieving practical convergence.}

\begin{figure}
    \centering    
\includegraphics[width=1\columnwidth]{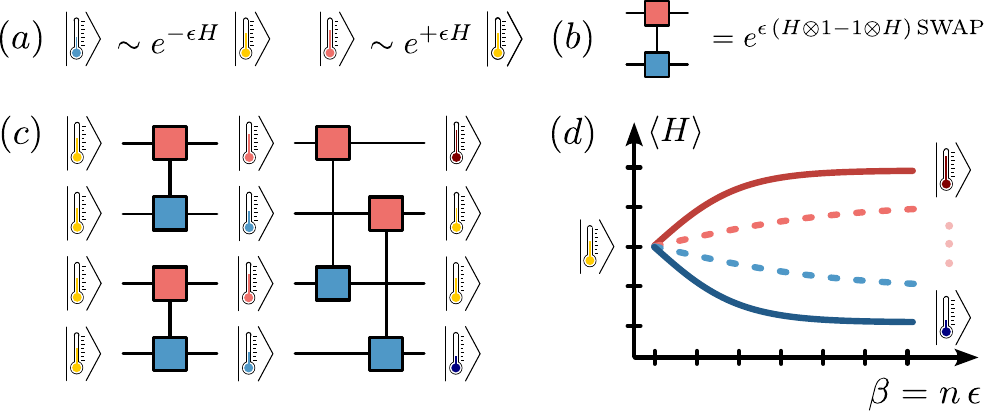}
    \caption{
    $(a)$ One small step of unitary imaginary time evolution $\propto \epsilon$ w.r.t.~the target Hamiltonian $H$
    ``heats'' or ``cools'' the state of an individual copy. $(b)$ We employ a unitary gate that can {approximately evolve} simultaneously a pair of copies forward and backward in imaginary time when applied to a pair that is symmetric under SWAP. $(c)$ The circuit illustrates how identically initialized copies can be re-used to collectively propagate multiple steps forward and backward in imaginary time. (d) In an idealized implementation of our protocols, the first (last) copy evolves forward (backward) in imaginary time through $n$ steps of size $\epsilon$, corresponding to a total imaginary time $\beta=n\epsilon$. As $\beta$ increases, the energy expectation value of the first (last) copy approaches the minimum (maximum) energy. The solid curves represent the first and last copies, while the dashed curves schematically represent intermediate copies that undergo fewer imaginary-time steps.}
\label{fig:opening}
\end{figure}

{Here we consider a deterministic approach based on quantum imaginary time evolution. Given a Hamiltonian $H$ and imaginary time $\beta$, the normalized imaginary time evolved pure state, represented in Fig.~\ref{fig:opening}(a), is 
\begin{equation}\label{pureQITE}
\ket{\phi_\beta} = \frac{e^{-\beta H}\ket{\phi}}{\sqrt{\bra{\phi} e^{-2\beta H} \ket{\phi}}},
\end{equation}}

Realizing such dynamics is attractive for several reasons. First, if $\ket{\phi}$ has non-zero overlap with the ground state, imaginary-time evolution drives it toward the ground state exponentially fast in $\beta$, and in certain cases this provides the fastest route to the ground state \cite{suzuki2025grover}; Second, if $\ket{\phi}$ is initialized to be a maximally entangled state with an auxiliary system on which $H$ has no support, then the imaginary time evolution, ignoring the auxiliary system, will produce the Gibbs thermal state proportional to $e^{-2 \beta H}$, giving direct access to finite temperature physics; Third, in many-body physics and quantum field theory, imaginary time turns unitary dynamics into a Euclidean theory, and Euclidean path integrals are a standard tool for defining states and studying dynamics under their influence~\cite{ginsparg1988applied, calabrese2007quantum}. In particular, in holographic settings \cite{maldacena1999large}, Euclidean path integrals are a way to explore states of key importance, such as ones dual to black holes and multi-boundary wormholes \cite{maldacena2003eternal, maldacena2004wormholes, balasubramanian2014multiboundary, brown2023quantum, nezami2021quantum, schuster2022many}.

Motivated by these connections, imaginary time evolution has become a central target for quantum state preparation protocols \cite{mcardle2019variational, QITE2020, mao2023measurement, huang2023efficient, hejazi2024adiabatic, gluza2024double, zander2025role, DBAC2025, suzuki2025double, robbiati2024double, gluza2024doubleDiag}. Existing approaches span variational formulations, measurement-assisted updates, and fully unitary circuit constructions based on structured transformations. While these methods have led to substantial progress, they come with characteristic tradeoffs, including ansatz dependency and trainability issues, measurement overhead and locality requirements, and nontrivial compilation or resource assumptions.

In this work, we construct explicit deterministic circuits that output an approximation of the imaginary time evolved state of \eqref{pureQITE} and can be used for ground state preparation. Our work is most closely related to recent algorithms based on the so-called \textit{double bracket flow} \cite{gluza2024double, zander2025role, DBAC2025, suzuki2025double, robbiati2024double, gluza2024doubleDiag}. 

The basic ingredients are the ability to store multiple copies of the system, and implement real-time evolution under $H$ on each copy and under the SWAP operator between the copies. The latter can be implemented using a controlled-SWAP mediated by a single auxiliary qubit. 
With these abilities, one can access an effective unitary that has the same effect to first order as the imaginary time evolution operator, $e^{\e (H\otimes 1- 1\otimes H)}$, when acting on two copies of a state (Fig.~\ref{fig:opening}(b)); this effective unitary "heats" one copy and "cools" the other, each approximately according to Eq. \eqref{pureQITE}. Then, more copies can be used to approximate larger imaginary times, as schematically illustrated in Fig. \ref{fig:opening}(c).

We analyze two circuit families, whose width is the number of copies, that show polynomial convergence with their depth to Eq.~\eqref{pureQITE}, and hence also to the ground state. 
The \textit{tree circuit} is provably convergent but requires a width that grows exponentially with depth, while the \textit{hedge circuit} is heuristic yet appears to achieve comparable performance with only polynomial width. We further show how post-selection can accelerate convergence with reasonable success probability, and we illustrate, through an explicit example, how one can trade circuit volume for measurements in the ground state preparation setting. Finally, we outline an implementation pathway relevant to near-term platforms, where multi-copy registers and SWAP-mediated interactions are natural, and we discuss how these hybrid analog–digital constructions can complement other state-preparation schemes when additional precision is needed.

\section{The protocol}

Under imaginary time evolution, the initial pure density matrix $\ket{\phi}\bra{\phi}$ remains pure and transforms to the density matrix,
\begin{equation} \label{betaState}
    \phi_\beta = \frac{e^{\beta H}\ket{\phi}\bra{\phi}e^{\beta H}}{\braket{\phi|e^{2\beta H}|\phi}}.
\end{equation}
The equation of motion that $\phi_\beta$ satisfies is given by,
\begin{equation} \label{ImEoM}
\begin{split}
    \frac{d\phi_\beta}{d\beta} & = H\phi_\beta + \phi_\beta H -2\phi_\beta \Tr(H\phi_\beta) 
    \\
    & = -[[\phi_\beta,H],\phi_\beta]
\end{split}
\end{equation}
where the fact that $\phi_\beta=\phi_\beta^2$, as the state is pure, was used.
The above dynamics is in the form of the so-called double-bracket flow, and describes the unitary evolution under the state-dependent Hamiltonian, $-i[\phi_{\beta},H]$.\footnote{The Hamiltonian here is the Hermitian operator $-i[\phi_{\beta},H]$.}

The first step of the protocol is to build unitary gates that, using additional copies, approximate the imaginary time evolution. These gates will be the building blocks of the circuits we discuss in the next sections.
We first note the following observation. Using the swap generator on a product state gives
\begin{equation}
\begin{split}
    \Tr_2(& e^{i \sqrt{\e} S}  \rho_1 \otimes  \rho_2 e^{-i \sqrt{\e} S}) 
    \\
    & = \rho_1 + i \sqrt{\e} [\rho_2,\rho_1]+\epsilon(\rho_2-\rho_1)+O(\e \sqrt{\e}),
    \label{SwapEvo}
\end{split}
\end{equation}
where $S$ is the SWAP operator between systems $1$ and $2$.
Equation \eqref{SwapEvo}  (derived together with Eqs. \eqref{SwapEvoForward} and \eqref{SwapEvoBackward} in Appendix \ref{SwapIdentitiesAppendix})  is the density matrix exponentiation trick, used to evolve $\rho_1$ according to a Hamiltonian that equals $\rho_2$ \cite{lloyd2014quantum, kimmel2017hamiltonian, kjaergaard2022demonstration, wei2023realizing, son2025quantum}.  

Consider now that $\rho_2$ is a copy of the pure state $\rho_1$, but evolved with the Hamiltonian $H$ for a time $\sqrt{\e}$, \ie $\rho_2 = e^{-i \sqrt{\e} H} \rho_1 e^{i \sqrt{\e} H} $.
Equation \eqref{SwapEvo} then becomes
\begin{equation}
\begin{split}
    \Tr_2(& e^{i \sqrt{\e} S}  \rho_1 \otimes  \rho_2 e^{-i \sqrt{\e} S})  
    \\
    & = \rho_1 + i \sqrt{\e}[\rho_1 - i\sqrt{\e}[H,\rho_1],\rho_1]+O(\e \sqrt{\e}) 
    \\
    & = \rho_1 - \e [[\rho_1,H],\rho_1] + O(\e \sqrt{\e}),
\end{split}
\label{SwapEvoForward}
\end{equation}
and 
\begin{equation}
\begin{split}
    \Tr_1(e^{i \sqrt{\e} S} & \rho_1 \otimes \rho_2 e^{-i \sqrt{\e} S})  
    \\
    & = \rho_2 + \e [[\rho_2,H],\rho_2] + O(\e \sqrt{\e}).
\end{split}
\label{SwapEvoBackward}
\end{equation}
According to Eq.~\eqref{ImEoM}, the last two equations show that to first order in $\e$, $\rho_1$ and $\rho_2$ evolve forward and backward in imaginary time, respectively. We note that both of them stay pure to first order. 

Based on this, we now define a gate $W$ acting on two identical systems, which approximates an $\e$ imaginary time step, as a $\sqrt{\e}$ real-time evolution on the second system, and a subsequent action on both systems with $e^{i \sqrt{\e} S}$, \ie
\begin{equation}
    W = e^{i \sqrt{\e} S} e^{-i\sqrt{\e} \, \boldsymbol{1} \otimes H}.
\end{equation}

Let us define
\begin{equation}
    \ket{a+i b} = \frac{1}{\sqrt{\braket{\phi|e^{2a H}|\phi}}}e^{(a-ib)H}\ket{\phi}.
\end{equation}
In this notation, we obtain that
\begin{equation}
    W\ket{a+i b} \otimes\ket{a+i b} \approx \ket{a+\e+i b} \otimes\ket{a-\e+i (b+\sqrt{\e})},
\end{equation}
where the trace distance to the ideal state is
\begin{equation}\label{traceNormDis}
    \frac{1}{2}\|\phi_{z+\e} \otimes \phi_{z-\e+i\sqrt{\e}} - W \phi_z \otimes \phi_z W^\dagger\|_1 =  D_z(H) \e^{3/2} + O(\e^{2}).
\end{equation}
Here, $z \equiv a+ib$, and $\phi_z = \ket{z}\bra{z}$. $D_z(H)$ depends on the energy probability distribution at the state $\phi_z$ and equals zero when $\phi_z$ is an eigenstate of $H$; see Appendix \ref{ElementaryGateErrors} for the expression for $D_z(H)$, as well as the derivation of Eqs. \eqref{traceNormDis}, \eqref{traceNormDisV5}, and \eqref{traceNormDisVtilde}.

The above gate, $W$, has the advantage of being generated by forward real-time evolution alone. 
For scenarios where backward evolution in real time becomes possible, we construct a simpler gate,
\begin{equation}\label{Utwo}
    U=e^{i\frac{\pi}{4}S}e^{-i \e K}e^{-i\frac{\pi}{4}S} = e^{ -\e K S},
\end{equation}
where $K = H \otimes \mathbf{1} - \mathbf{1} \otimes H$. 
To first order in $\e$, $U$ has the same action on two copies of a pure state as the imaginary time evolution according to the Hamiltonian $K$.
The trace distance is given by,
\begin{equation}\label{traceNormDisV5}
    \frac{1}{2}\|\phi_{z-\e} \otimes \phi_{z + \e} - U \phi_z \otimes \phi_z U^\dagger\|_1 =  \sqrt{\text{Var}_z(K^2)} \e^{2} + O(\e^3),
\end{equation}
where $\text{Var}_z(K^2) = \Tr(K^4 \phi_z\otimes\phi_z) - \Tr(K^2 \phi_z\otimes\phi_z)^2$, is the variance of $K^2 = (H \otimes 1-1 \otimes H)^2$.
We note that the above gates are not unique, and using the generators $S, \, H \otimes \mathbf{1},$ and $\mathbf{1} \otimes H$, one can construct other unitaries that approximate Eq. \eqref{ImEoM} to first order. 

{As a final example, we construct the gate 
\begin{equation}\label{newV}
    V = e^{i\frac{\pi}{4}S}e^{-i 2\e H \otimes 1}e^{-i\frac{\pi}{4}S}
\end{equation}
which satisfies
\begin{equation}\label{traceNormDisVtilde}
\begin{split}
    \frac{1}{2}\|\phi_{z-\e+i \e} \otimes \phi_{z + \e+i \e} & -  V \phi_z \otimes \phi_z  V^\dagger\|_1 
    \\
    & =  \sqrt{\text{Var}_z(K^2)} \e^{2} + O(\e^3),
\end{split}
\end{equation}
and has the advantage of using only forward real-time evolution and an error that is proportional to $\e^2$.}

We have described gates that approximate an incremental imaginary time evolution, given two copies of the same pure state. One copy evolves forward, while the other evolves backward. After this increment, the two states are no longer the same. In the following section, we describe how circuits composed of the above gates, acting on more copies of the initial state, can provide an approximate evolution to larger imaginary times, and therefore can also be used for ground state preparation.

\subsection{Compilation guidelines}\label{secComp}
Let us start by describing a heuristic guideline used in the protocols.
The plan is to use more copies and to continue acting with $U$ (or $V$) between the different copies, given that they are approximately in the same state. By using enough copies such that $n$ $U$-gates acted on the first copy, and by choosing $\e = \frac{\beta}{n}$, we can obtain a final state in which the first copy is approximately evolved by $\beta$, as in Eq.\eqref{betaState}.

Let us define $U_{ij}$ (or $V_{ij}$) as the gates $U$($V$) acting on copies $i$ and $j$,
\begin{equation}\label{Uij}
    U_{ij}=e^{i\frac{\pi}{4}S_{ij}}e^{-i \e K_{ij}}e^{-i\frac{\pi}{4}S_{ij}} = e^{ -\e K_{ij}S_{ij}},
\end{equation}
where $K_{ij} = H_i - H_j$, with $H_i$ being the Hamiltonian $H$ acting on system $i$, and $S_{ij}$ the swap operator between system $i$ and $j$.

More schematically, the guideline is,
\begin{itemize}
    \item{1)} Prepare $2 m$ copies of the initial state, $\ket{0}^{\otimes 2 m}$, and a bookkeeping classical vector $\vec c=(0,...,0)$ of $2m$ dimensions,
    \item{ 2)} Find indices $i,j$, with $i<j$ of two identical elements in $\vec c$. Act on the state with $U_{ij}$, and add $+1$ to $c_i$, and $-1$ to $c_j$. For circuits composed out of $V_{ij}$ gates instead of $U_{ij}$, add $-1+i$ to $c_j$ instead of just $-1$. If there are elements with identical real and different imaginary values, one can adjust by doing real time evolution on the copy with the smaller imaginary value before using $V_{ij}$,
    \item{3)} Repeat until $c_1 =n$ to obtain an approximate $e^{n \e H}$ evolved state in the first copy. 
\end{itemize}
The restriction to equal bookkeeping entries ensures that, in the ideal bookkeeping picture, the two input copies are identical, which is the regime in which the two-copy gate implements the perturbative imaginary-time update.\footnote{Acting on copies with different entries of $\vec c$ is still a well-defined unitary operation. Such unbalanced gates may be useful as heuristic or variational additions, for example near the end of the circuit, but they are not part of the constructions analyzed here.}

As an illustration of the above guideline, using the $U_{ij}$ gates on $2m=4$ copies, we can generate the following process:
\begin{equation} \label{FourExample}
\begin{split}
    \ket{0,0,0,0} & \to \ket{\e,-\e,\e,-\e} \\
    &
    \to \ket{2 \e,0,0,-2 \e} 
    \\
    &\to \ket{2 \e,\e ,-\e,-2 \e},
\end{split}
\end{equation}
where we note that the actual state of the four copies along the circuit is an approximation of the process above. As two gates touched the first copy, its error to the ideal state $\ket{2 \e}$ will be roughly $O(2 \e^2)$. Taking $\e = \beta/2$, results in an $O(\beta^2/2)$ error, which is a factor of $1/2$ better than using just two copies to reach the same imaginary time.
More generally, as there are $n$ gates acting on the first copy, summing over the single gate errors, which according to Eq.~\eqref{traceNormDisV5} scale like $\e^2$, gives the expectation that the final error will scale as $O(\frac{1}{n})$.

Using this guideline, we now describe three protocols. We start with the tree-circuit (see Fig.~\ref{fig:CircuitsCombined}(a)): a circuit with a tree-like structure and a number of copies that is exponential in the number of evolution steps on the first copy. For this protocol, we derive a convergent upper bound on the error.

The second is the hedge-circuit (see Fig.~\ref{fig:CircuitsCombined}(b)): a heuristically more efficient protocol that requires a polynomial number of copies in the number of evolution steps on the first copy. In addition, we discuss how measurements and post-selection along the first two circuits can accelerate their convergence. 

Finally, in the third protocol, we describe how the circuit volume can be interchanged with the number of measurements needed to evaluate the final state. Specifically, we consider a one-layer circuit that prepares copies that are approximately thermal Gibbs states with a small inverse temperature, and use a swap trick to estimate observables in the ground state. 

To evaluate the output of the protocols, we define $\mathcal{F}_{\beta}$ as the fidelity between the state of the first copy and the exact imaginary time evolved state, \ie \footnote{We note that $\phi_\beta$, defined in Eq. \eqref{betaState}, is pure, and therefore the expression for the fidelity simplifies to the trace of the product of the two states.}
\begin{equation}\label{Fidelity}
    \mathcal{F}_{\beta} = \Tr(\phi_{-\beta} \tilde \phi_1^{(n)}),
\end{equation}
where $\tilde \phi_1^{(n)}$ is the final state of the first copy of the circuit, and the superscript denotes the number of evolution steps performed on it, which also controls the number of copies used in the circuit.
We also denote the fidelity with the ground state as
\begin{equation}
    \mathcal{F} = \Tr(\phi_{gs} \tilde \phi_1^{(n)}),
\end{equation} 
where $\phi_{gs}$ is the ground state.

To represent the circuits in diagrammatic notation, we introduce the following diagram for the $U_{ij}$ gates:
\[
U_{ij} \;=\;
\begin{tikzpicture}[baseline={(current bounding box.center)},
  x=0.8cm, y=0.55cm,
  wire/.style={very thick, line cap=round},
  link/.style={very thick, line cap=round},
  bdot/.style={circle, fill=blue!65, draw=none, minimum size=4.5pt, inner sep=0pt},
  rdot/.style={circle, fill=red!80,  draw=none, minimum size=4.5pt, inner sep=0pt},
  lab/.style={font=\small}
]
  \def\xL{0}
  \def\xR{1.4}
  \def\xg{0.7}

  \draw[wire] (\xL,0) -- (\xR,0);
  \draw[wire] (\xL,-1) -- (\xR,-1);

  \node[lab,anchor=east] at (\xg,0.5) {$i$};
  \node[lab,anchor=east] at (\xg,-1.5) {$j$};

  \draw[link] (\xg,0) -- (\xg,-1);
  \node[bdot] at (\xg,0) {};
  \node[rdot] at (\xg,-1) {};
\end{tikzpicture}
\]
where the $i$ and $j$ indices in the diagram mark the lines representing the $i$ and $j$ copies. The upper copy, coupled using the blue dot, evolves approximately by an imaginary time $-\e$, whereas the lower copy evolves by $\e$. We note that in the following circuits, whether one wants the final state to approximate a plus or minus $\beta$ evolution can be altered, for example, by changing the sign of $S_{ij}$ or $K_{ij}$ in Eq.~\eqref{Uij}.   

\subsection{The tree circuit} \label{exponential}
The tree-circuit requires $2^{n}$ copies to advance the first copy $n$ steps. The circuit is organized such that the gates are always acting on two copies with an exactly identical state.
One way to construct the circuit is with $n$ layers, where each layer has half the number of gates as the previous layer. After each layer, only half the copies of the previous layer are used - those that were approximately time evolved in the same direction. 

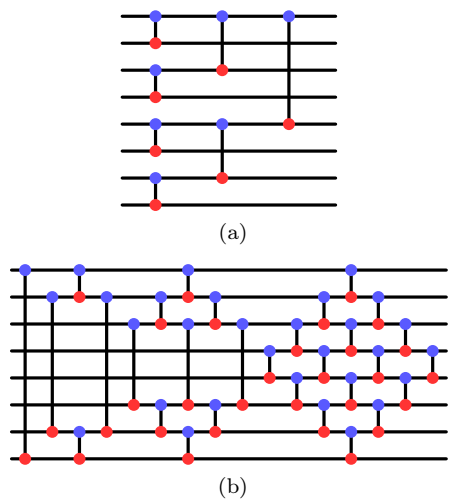
\begin{figure}[b]
\centering

\subfloat[]{\label{fig:TreeCircuit}
\begin{tikzpicture}[scale = 0.65,
  x=1.35cm, y=0.55cm,
  wire/.style={very thick, line cap=round},
  link/.style={very thick, line cap=round},
  bdot/.style={circle, fill=blue!65, draw=none, minimum size=4.5pt, inner sep=0pt},
  rdot/.style={circle, fill=red!80,  draw=none, minimum size=4.5pt, inner sep=0pt}
]

\def\xL{0}
\def\xR{3.2}
\foreach \i in {0,...,7} {
  \draw[wire] (\xL,-\i) -- (\xR,-\i);
}

\def\xa{0.5}
\foreach \top/\bot in {0/1, 2/3, 4/5, 6/7} {
  \draw[link] (\xa,-\top) -- (\xa,-\bot);
  \node[bdot] at (\xa,-\top) {};
  \node[rdot] at (\xa,-\bot) {};
}

\def\xb{1.5}
\foreach \top/\bot in {0/2, 4/6} {
  \draw[link] (\xb,-\top) -- (\xb,-\bot);
  \node[bdot] at (\xb,-\top) {};
  \node[rdot] at (\xb,-\bot) {};
}

\def\xc{2.5}
\draw[link] (\xc,0) -- (\xc,-4);
\node[bdot] at (\xc,0) {};
\node[rdot] at (\xc,-4) {};

\end{tikzpicture}}\\[0.6em]
\subfloat[]{\label{fig:HedgeCircuit}

\begin{tikzpicture}[scale=0.65,
  x=0.55cm, y=0.55cm,
  wire/.style={very thick, line cap=round},
  link/.style={very thick, line cap=round},
  bdot/.style={circle, fill=blue!65, draw=none, minimum size=4.5pt, inner sep=0pt},
  rdot/.style={circle, fill=red!80,  draw=none, minimum size=4.5pt, inner sep=0pt}
]

\newcommand{\vlink}[3]{%
  \draw[link] (#1,-#2) -- (#1,-#3);
  \node[bdot] at (#1,-#2) {};
  \node[rdot] at (#1,-#3) {};
}

\def\xL{-0.5}
\def\xR{15.5}
\foreach \i in {0,...,7} {
  \draw[wire] (\xL,-\i) -- (\xR,-\i);
}

\vlink{0}{0}{7}
\vlink{1}{1}{6}
\vlink{2}{0}{1}
\vlink{2}{6}{7}
\vlink{3}{1}{6}
\vlink{4}{2}{5}
\vlink{5}{1}{2}
\vlink{5}{5}{6}
\vlink{6}{0}{1}
\vlink{6}{2}{5}
\vlink{6}{6}{7}
\vlink{7}{1}{2}
\vlink{7}{5}{6}
\vlink{8}{2}{5}
\vlink{9}{3}{4}
\vlink{10}{2}{3}
\vlink{10}{4}{5}
\vlink{11}{1}{2}
\vlink{11}{3}{4}
\vlink{11}{5}{6}
\vlink{12}{0}{1}
\vlink{12}{2}{3}
\vlink{12}{4}{5}
\vlink{12}{6}{7}
\vlink{13}{1}{2}
\vlink{13}{3}{4}
\vlink{13}{5}{6}
\vlink{14}{2}{3}
\vlink{14}{4}{5}
\vlink{15}{3}{4}

\end{tikzpicture}}

\caption{(a) The tree circuit and (b) the hedge circuit, each for eight copies. To advance the first copy $n$ steps, the tree circuit uses $2^n$ copies and $2^n-1$ gates, while the hedge circuit uses $2n$ copies and $\sim n^3$ gates.}
\label{fig:CircuitsCombined}
\end{figure}

More precisely, the first layer is composed of the product of gates, $\Pi_{i=0}^{(2^n/2)-1}  U_{2 i+1,2i+2}$. The second layer is $\Pi_{i=0}^{(2^n/4)-1}  U_{4 i+1,4i+3}$, and in general, layer $l$ is given by the gates $\Pi_{i=0}^{2^{n-l}-1}  U_{2^l i+1,2^l i+1+2^{l-1}}$.
This produces a tree-like structure, as in the $8$-copy example in Fig.~\ref{fig:TreeCircuit}.

Under this circuit, the bookkeeping vector $\vec c$ follows the trajectory $(0,...,0)\to(-1,1,...,-1,1)\to(-2,1,0,1,...,-2,1,0,1)\to ... \,$.\footnote{With our sign convention, the tree circuit described here takes the first copy to the bookkeeping value $-n$. This choice is arbitrary: one could instead follow the opposite output branch at each layer, which would take, for example, the last copy to $+n$. One can also follow both branches simultaneously, as in the four-copy example of Eq.~\eqref{FourExample}, so that the final bookkeeping values of the first and last copies are $(-n,n)$. Although these two output copies are generally correlated, each individual reduced state follows the recurrence of Eq.~\eqref{TreeRecurRel}, with the sign of $\epsilon$ in $U_{12}$ determined by the branch being followed. No measurement or reinitialization is involved in this deterministic construction; these operations are only introduced later in the post-selected variant.} In layer $l+1$, the full quantum circuit acts on the $2^{n-l}$ copies whose corresponding entries in $\vec c$ are equal to $-l$. Since we are ultimately interested in the final reduced state of the first copy, it is useful, for analytical purposes and for classical simulations of the protocol, to focus on its trajectory through the circuit. As we explain below, the tree structure implies that, for this purpose, one does not need to keep track of the full many-copy state, but only of an effective two-copy evolution, even though the actual quantum circuit acts on the full many-copy system.

Let $\tilde \phi^{(l)}$ denote the state of all $2^n$ copies after layer $l$, and define the reduced state of the first copy as $\tilde \phi_1^{(l)}=\Tr_{\bar 1}(\tilde \phi^{(l)})$, where $\Tr_{\bar 1}$ denotes the trace over all copies except copy $1$. Along the trajectory of the first copy, layer $l+1$ couples copy $1$ to copy $1+2^l$ through the two-copy gate $U_{1,1+2^l}$. Before this gate is applied, these two copies have evolved through identical but disjoint subtrees. Hence their reduced states are identical, and their joint reduced state factorizes as $\tilde \phi_1^{(l)}\otimes \tilde \phi_1^{(l)}$. The update of the first copy is therefore obtained by applying $U_{1,1+2^l}$ to this two-copy state and tracing out copy $1+2^l$. Since the same two-copy gate is used at each layer, only on different copy labels, we can relabel copy $1+2^l$ as copy $2$ and write the evolution of $\tilde \phi_1^{(l)}$ as the following two-copy recurrence:
\begin{equation} \label{TreeRecurRel}
    \begin{split}
        \tilde \phi_1^{(l+1)} = \Tr_2(U_{12} \, \tilde \phi_1^{(l)}\otimes \tilde \phi_1^{(l)} U_{12}^\dagger).
    \end{split}
\end{equation}

\begin{figure*}[t]
    \centering
    \subfloat[]{\includegraphics[width=0.33\textwidth]{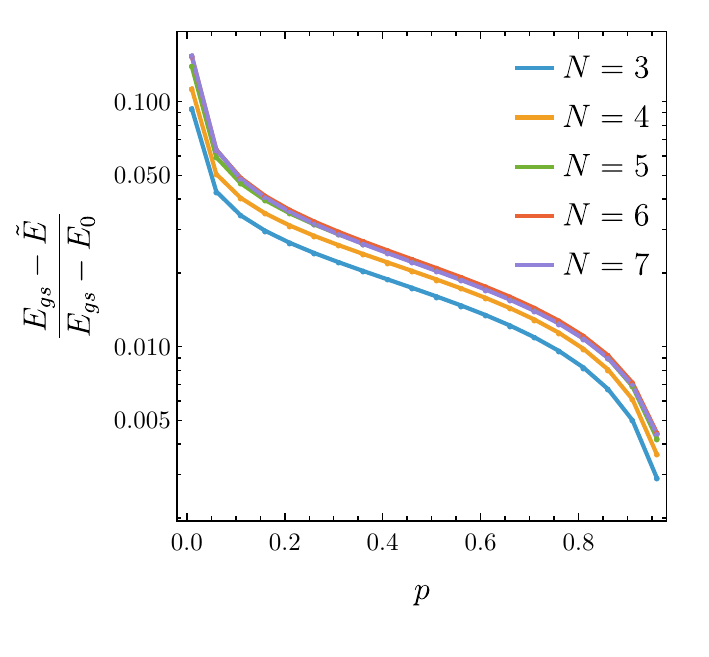}{}}\hfill
    \subfloat[]{\includegraphics[width=0.33\textwidth]{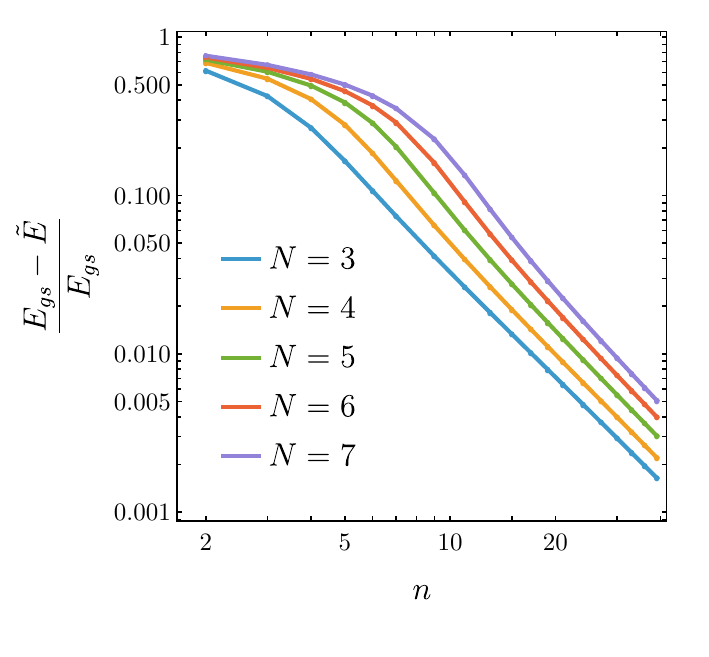}{}}\hfill
    \subfloat[]{\includegraphics[width=0.33\textwidth]{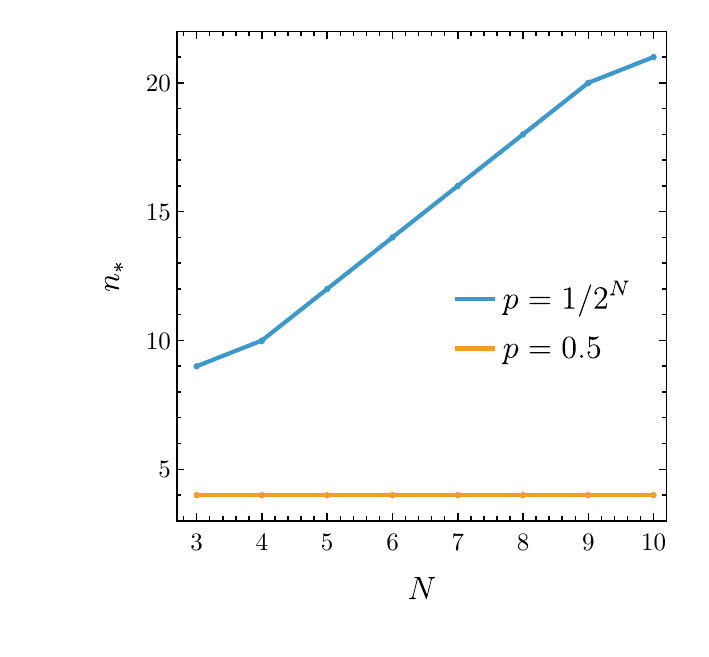}{}}
    \caption{Numerical results for the ground state preparation of the mixed field Ising Hamiltonian $H=-\sum_{i}( \sigma_z^{i}\sigma_z^{i+1}+\sigma_z^{i} + \sigma_x^{i})$ with periodic boundary conditions and $N$ spins, using the tree circuit in Sec.~\ref{exponential}. The initial pure state is defined as $\ket{\phi_0} = \sqrt{p}\ket{\phi_{gs}}+\sqrt{1-p}\ket{\phi_{\perp}}$, where $\ket{\phi_{\perp}}$ is a normalized equal weight superposition of all energy eigenvectors, except the ground state, $\ket{\phi_{gs}}$. The ground state energy is $E_{gs}$. All data points are given by minimizing the protocol's output energy, $\tilde E$, over $\e$.  $(a)$ Relative reduction in energy, $(E_{gs}-\tilde E)/(E_{gs}-E_0)$ as a function of $p$, for $n=10$ layers. Here, $E_0$ is the energy expectation value of the protocol's input state.  $(b)$ $(E_{gs}-\tilde E)/(E_{gs})$ as a function of the number of layers with $p=1/2^N$. 
    $(c)$ The minimal $n$ needed for $(E_{gs}-\tilde E)/(E_{gs})\leq 0.05$, $n_*$, as a function of $N$. Here, $p=1/2^N$ represents an input state with a random-like behavior, and $p=0.5$ represents an input state that has been prepared to be close to the ground state.}
    \label{fig:three_panels}
\end{figure*}

For the tree circuit, we derive in Appendix \ref{TreeErrorBound} the following upper bound:
\begin{equation}\label{boundExp}
\|\phi_{\beta} - \tilde \phi_1^{(n)}\|_1\leq \frac{\beta \, \sigma_{l_*}(K^2)}{2n \|H\| } (e^{4 \beta \|H\|}-1)+O\left(\frac{1}{n^2}\right),
\end{equation}
where $\tilde \phi_1^{(n)}$ is the final state of the first copy, and $\phi_{\beta}$ is the exact imaginary evolved state to $\beta$.
Not surprisingly, this upper bound has a similar form to the bound derived in Ref.~\cite{mcmahon2025equating}, where their Riemannian gradient descent-based algorithm also approximates the double bracket flow in Eq.~\eqref{ImEoM}. 
Using the triangle inequality, Eq. \eqref{boundExp} can be used to upper bound the trace distance with the ground state. 
\begin{equation} \label{boundGS}
    \|\phi_{gs} - \tilde \phi_1^{(n)}\|_1\leq \|\phi_{gs} - \phi_{\beta}\|_1+\|\phi_{\beta} - \tilde \phi_1^{(n)}\|_1.
\end{equation}
For systems for which $\|H\|$ is extensive in the number of degrees of freedom, Eqs.~\eqref{boundExp} {and \eqref{boundGS}} scale exponentially with the system size. 
However, as noted in Ref.~\cite{pervez2025riemannian} and seen in their numerical results for ground state preparation, the actual convergence for certain systems can be significantly faster. 

In Fig.~\ref{fig:three_panels}, we provide numerical results for the ground state preparation of the mixed field Ising model $H=-\sum_{i}( \sigma_z^{i}\sigma_z^{i+1}+\sigma_z^{i} + \sigma_x^{i})$ with periodic boundary conditions and $N$ spins. The results are obtained by minimizing the energy expectation value of the output state over $\e$, the step size. Although limited to a small system size, the data suggest that for this Hamiltonian (whose operator norm, $\|H\|$, is extensive in $N$), the convergence rate to the ground state depends mainly on the energy gap and the initial state's overlap with the ground state. For a fixed overlap with the ground state, the results are insensitive to the system size, and when the overlap is set to be $1/2^N$, emulating the overlap of a random state, the results depend polynomially on the size. In addition, the protocol's performance improves as the initial state gets closer to the ground state, pointing to the usefulness of this protocol in accompanying other ground state preparation methods; see, for instance, Refs.~\cite{ hauke2020perspectives, cerezo2021variational, tilly2022variational, zhou2020quantum, ebadi2022quantum, bombieri2025quantum}.

Nevertheless, although for certain systems the protocol could converge much faster than the upper bound of Eq.~\eqref{boundExp}, there is still an exponential overhead in the tree-like structure of the protocol. This limitation motivates the more compact circuit introduced in the next section.

\subsection{The hedge circuit}\label{hedge}

The previous tree circuit uses $2^n$ copies to approximately evolve the first copy to  $\beta= -n \e$. This is done by acting with the $U_{ij}$ gate on the first copy $n$ times. However, following the heuristic guideline introduced at the beginning of Sec.~\ref{secComp}, it is possible to act $n$ times on the first copy using only $2n$ copies, before there are no more equal elements in the bookkeeping vector $\vec c$. The final state of $\vec c$, for which there are no more equal elements is given, up to permutations of the elements, by $\vec c_f = (-n,...,-1,1,...,n)$, and the circuit achieving this is not unique.

One particular circuit configuration that achieves the above protocol is composed of diamond shape blocks where each block acts on two extra copies compared to the previous block, as in Fig.~\ref{fig:HedgeCircuit}. 
To be more precise, the algorithm is constructed as follows.
We first define the list of copies $v^{kp}=(k,...,p,2n-p+1,...,2n-k+1)$, where we denote the $i$-th element of $v^{kp}$ from left to right as $v^{kp}_i$. Then, we define the circuit layer $L^k_p =\prod_{i=1}^{p-k+1} U_{v^{kp}_{2i-1},v^{kp}_{2i}}$. From this layer, we build the blocks $B_i= L^{i}_iL^{i-1}_i...L^{1}_i...L_i^{i-1}L^{i}_i$, \ie the upper index starts from $i$, decreases by one in the next term of the product, until it reaches $1$, and then increases by one in the next term of the product, until it reaches $i$ again. Finally, the full circuit on the $2n$ copies is given by $B_n...B_2B_1$.

As each block $B_i$ contains $i^2$ gates, reaching the $\vec c_f$ target state requires $M =  n(n+1)(2n+1)/6 \sim n^3$ actions of $U_{ij}$, \footnote{This is the optimal gate count regardless of the circuit structure. To see this, let us represent the bookkeeping vector after the $m$-th use of $U_{ij}$ as $\vec c^{(m)}$. 
Consider the following function $S(\vec c^{(m)}) = (c^{(m)}_1)^2+(c^{(m)}_2)^2+...$, where $c^{(m)}_i$ are the elements of $\vec c^{(m)}$. As $U_{ij}$ only acts on two elements if they are equal, and it adds $1$ to one of them and $-1$ to the other, we understand that $S(\vec c^{(m+1)})-S(\vec c^{(m)}) = 2$. Therefore, $S(\vec c^{(m)}) = 2 m$, and the minimal number $M$ to reach a state is given by $S(\vec c^{(m)})/2$. For the state $\vec c_f=(-n,...,-1,1,...,n)$, we obtain the claimed result.} which gives an exponential improvement of circuit complexity over the tree-circuit.

The analytical evaluation of the hedge-circuit's convergence is left for future work. Here, we focus on providing numerical support by computing the output's (in)fidelity with the exact imaginary time evolved state according to \eqref{Fidelity}, where we expect it to be $\leq O(\frac{1}{n})$.  In addition, for the ground state preparation, we minimize the energy expectation value of the protocol's output over the step size $\e$, and compute the difference with the exact ground state energy. 
Given the limitation of classical computers, we focus on a single qubit Hamiltonian, $H=\sigma_z$.

\begin{figure*}[t]
\centering

\subfloat[]{\label{fig:fidelities_chi400}%
\includegraphics[width=0.5\textwidth]{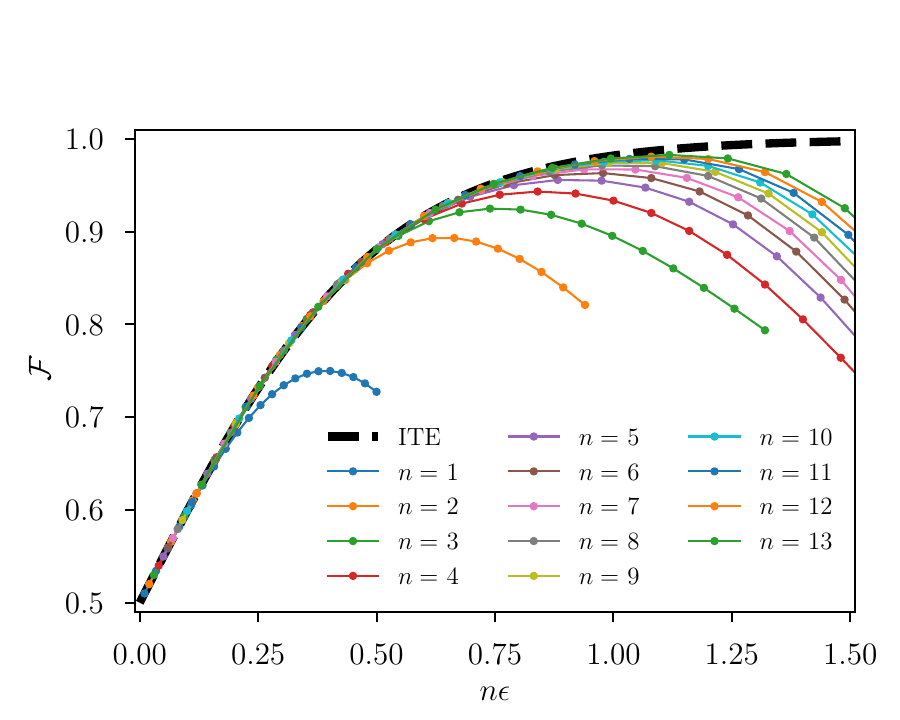}}
\hfill
\subfloat[]{\label{fig:convergence_loglog}%
\includegraphics[width=0.5\textwidth]{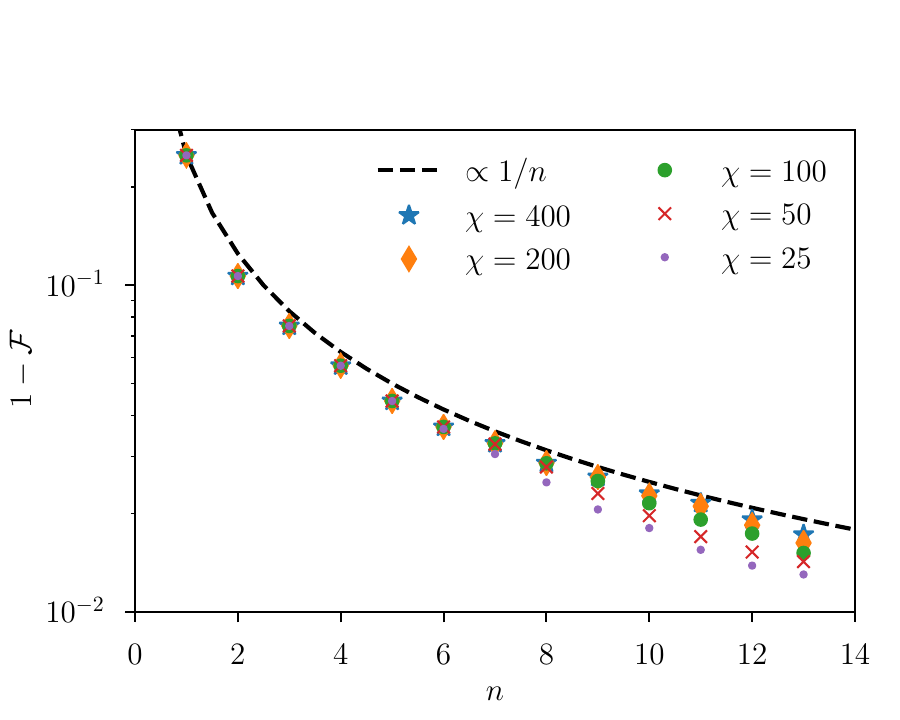}}

\vspace{0.5em}

\hspace*{-0.02\textwidth}
\subfloat[]{\label{fig:EnrgyDiffWithProj}%
\includegraphics[width=0.465\textwidth]{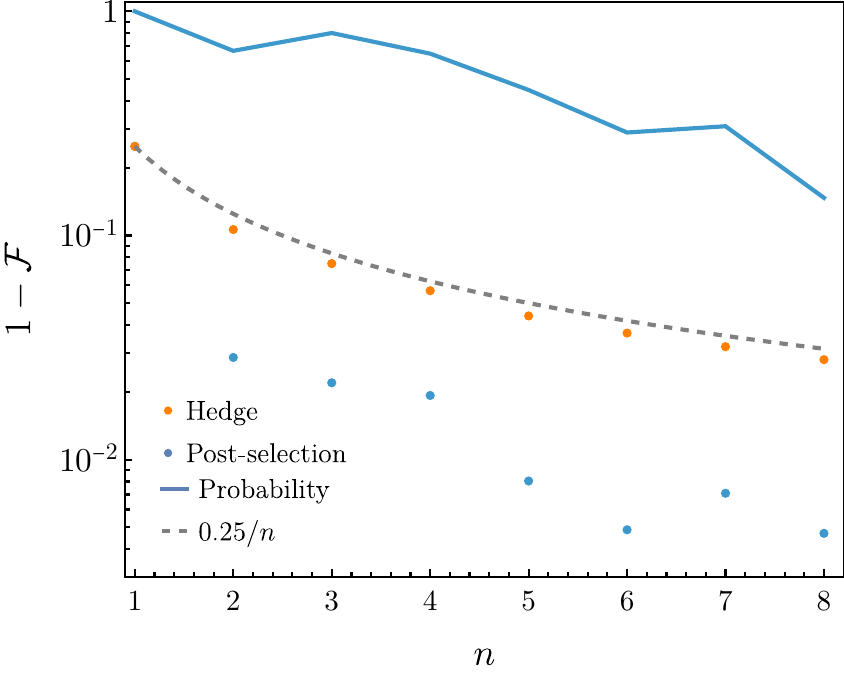}}
\hspace*{0.01\textwidth}
\subfloat[]{\label{fig:convergence_loglogED}%
\includegraphics[width=0.493\textwidth]{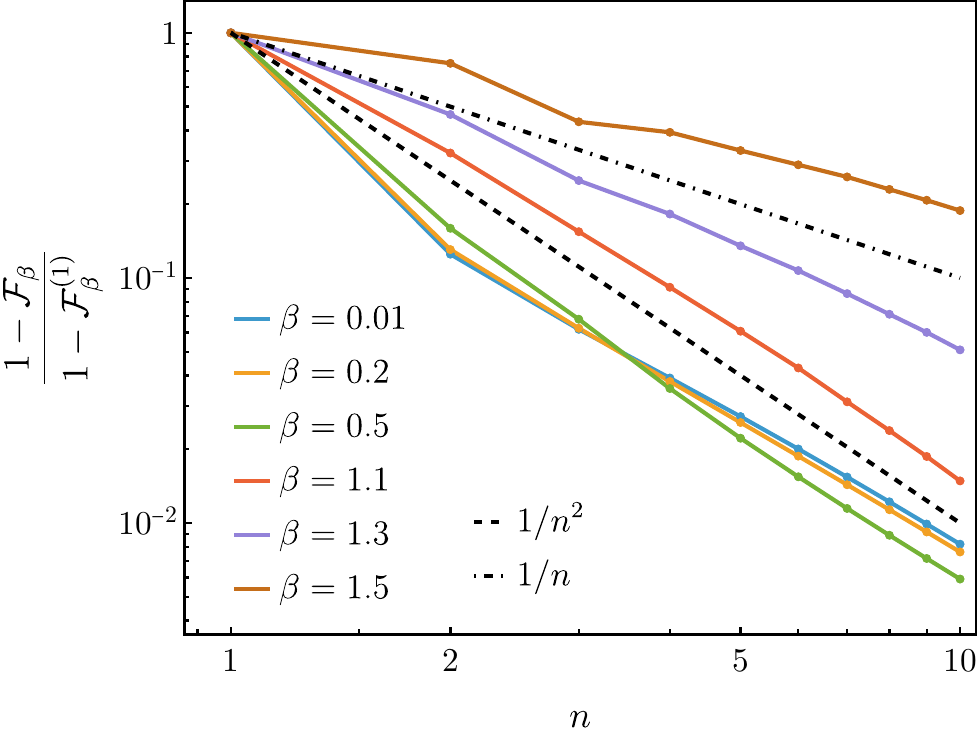}}

\caption{Simulations of the protocol with the hedge circuit in Sec.~\ref{hedge} with $H=\sigma_z$. The initial state is given by $2n$ copies of the $-\sigma_x$ ground state.  (a) MPS simulation using bond dimension $\chi =400$ for the fidelity with the ground state, as a function of $n \e$. The dashed curve represents the exact imaginary time evolution with imaginary time $n\e$. (b) MPS simulation for the infidelity with the exact ground state as a function of the number of copies, where the protocol is optimized over the step parameter $\e$ to give minimal energy. $\chi$ is the bond dimension. (c) Comparison of the hedge circuit performance for ground state preparation, given by the distance to the ground state energy, with and without post-selection. The solid line is the total probability of the post-selection. Both protocols are optimized over the step parameter $\e$ to give minimal energy. (d) Infidelity with the exact imaginary evolved state as a function of $n$, normalized to $1$ at $n=1$. The infidelity at $n=1$ for the different increasing values of $\beta$ is $1-\mathcal{F}_\beta^{(1)} = (0.39 \cdot 10^{-3},0.26,0.27,0.50,0.25,0.58)$.}
\label{fig:four_grid}
\end{figure*}

For ground state preparation, we observe a scaling of $1/n$ for the infidelity between the protocol's output and the exact ground state. This can be seen in the MPS-based numerical results presented in Figs.~\ref{fig:fidelities_chi400} and ~\ref{fig:convergence_loglog}. Similar scaling can be seen in Fig.~\ref{fig:EnrgyDiffWithProj} using exact diagonalization. 

For imaginary time evolution, as can be seen in the $\log-\log$ plot of Fig.~\ref{fig:convergence_loglogED}, the infidelity with the exact state scales as a power law in regions where $n$ is large enough (such that the single gate error in Eq.~\eqref{traceNormDisV5} is small). Interestingly, for certain choices of $\beta$, the infidelity scales as $1/n^2$, as opposed to $1/n$, however, results for a larger number of copies are required to be conclusive. In addition, the rate of convergence slows as $\beta$ increases; however, for large $\beta$, $\phi_\beta$ is close to the ground state, and therefore, the more efficient ground state preparation scheme can be used instead.

{\it Post-selection protocol.} Although the target state of the protocol is pure, the results of Fig.~\ref{fig:convergence_loglog} suggest that there is significant entanglement between the copies along the circuit, and that reducing it with a finite bond dimension speeds up the convergence. This brings us to study a version of the circuit that includes post-selection. In the post-selected version, as we keep track of the bookkeeping vector $\vec c$, every time a copy approximately reaches its initial state, \ie when its corresponding vector element equals $0$, the copy is measured and post-selected to the initial state $\ket{0}$.

In Fig.~\ref{fig:EnrgyDiffWithProj}, we observe that the post-selected version provides substantial speedup in the ground state convergence, at the cost of making the protocol reasonably probabilistic. This can be especially useful in resource-limited settings, such as a limited number of copies or high-fidelity gates.

\subsection{Interchanging circuit volume with number of measurements} \label{interchange}
The final protocol demonstrates a way to interchange the number of copies and gates with the number of measurements needed in order to evaluate ground state observables.  The first modification is that instead of approximating an imaginary time evolution on a generic pure state, half of the maximally entangled states are evolved, and therefore the target state of the protocol is the actual thermal state $e^{-\beta H}/\Tr(e^{-\beta H})$. Then, with several initial copies, one can end the protocols discussed thus far, with one copy evolved in imaginary time as much as possible, or alternatively, end the protocols with more copies evolved to shorter times. 

Having more copies evolved to inverse temperature $\e$, say $n$, one can perform \textit{virtual cooling}~\cite{QVC2019} (see also Ref.~\cite{huggins2021virtual} for related ideas) to probe $n \e$ temperatures, and approach exponentially fast to the ground state. This is performed using the swap trick, summarized in the following relation,
\begin{equation} \label{swaptrick}
    \frac{\Tr(x_1 S_{n-1 \, n}...S_{23}S_{12}\rho^{\otimes{n}})}{\Tr(S_{n-1 \, n}...S_{23}S_{12}\rho^{\otimes{n}})} = \frac{\Tr(x_1 \rho^n)}{\Tr(\rho^n)}.
\end{equation}
If $\rho \propto e^{-\e H }$, the right-most term in the above equation is the expectation value under the thermal state with inverse temperature, $\beta= n\e$. The cyclic-SWAP identity in Eq.~\eqref{swaptrick} is the same replica trick that underlies multi-copy measurements of R\'enyi entropies and entanglement spectra, as discussed for example in
Refs. ~\cite{ QVC2019,pichler2013thermal}. From an experimental perspective, in 
the present protocol, 
the SWAP operations play two crucial and complementary roles: they first appear in the preparation of approximate
imaginary-time-evolved copies and then in the final replica measurement used for virtual cooling.

Interestingly, the error in the protocol's output $\tilde \rho$ from the exact thermal state limits the usage of the swap trick to probe generic inverse temperatures. However, as we will show below, it works for the ground state estimation, where $n \e \to \infty$.
To demonstrate the method, we shall focus on the case in which we minimize the number of gates and maximize the number of copies to be used in the swap trick. Let us start with $2 n$ copies. Consider a circuit composed of a single layer of the form $U_{2n,2n-1}...U_{5,6}U_{3,4}U_{1,2}$. Having used $n$ unitaries, the state has an error $ \sim n \e^2$ from the ideal imaginary evolved state $(\e,-\e,\e,-\e,...)$. Tracing over the odd copies results in the state $(-\e,-\e,...)$ without increasing the error. Each of the $2n$ copies is set to be composed out of $2$ internal copies, with the initial state being the maximally entangled state. 

The Hamiltonian, used in $U_{ij}$, according to which the state is approximately imaginary evolved, is $H=H_{int}\otimes 1$. Therefore, the current state $(-\e,-\e,...)$ approximates the state,
\begin{equation}
\begin{split}
    \bigotimes_{i=1}^n \sum_{i,j} & \frac{e^{-\e H}\ket{ii}\bra{jj}e^{-\e H}}{\Tr(e^{-2\e H})} =
    \\
    & \bigotimes_{i=1}^n \sum_{i,j}\frac{e^{-\e H_{int}}\otimes 1\ket{ii}\bra{jj}e^{-\e H_{int}}\otimes 1}{\Tr(e^{-2\e H_{int}})},
\end{split}
\end{equation}
where tracing over the second internal copy of each of the copies produces a product of thermal states with inverse temperature $2\e$,
\begin{equation}
    \bigotimes_{i=1}^n \frac{e^{-2\e H_{int}}}{\Tr(e^{-2\e H_{int}})},
\end{equation}
while the bound on the error remains $\sim O(n\e^2)$.

Having at our disposal an approximate state of $n$ copies of the thermal state, we can use the swap operator on the internal copies to obtain the expectation value.
This means that it is possible to obtain expectation values under the operator $\tilde \rho^n=\rho^n+O(n \e ^2)$, where
\begin{equation}
    \tilde\rho^n= \rho^n+O(n \e^2)=\frac{e^{-2 n \e H_{int}}}{\Tr(e^{-2  \e H_{int}})^n}+ O(n \e^2).
\end{equation}

Although $\rho^n$ can be used to obtain information on the spectrum of $H_{int}$, the actual thermal state with inverse temperature $n \e$ is given by $\rho^n/\Tr(\rho^n)$, which unfortunately, $\tilde \rho^n/\Tr(\tilde \rho^n)$ do not necessarily converge to. Nevertheless, we shall now show that with large $n$, $\tilde \rho^n/\Tr(\tilde \rho^n)$ converges to an eigenstate of $H_{int}$, and specifically, if $\e$ is small enough, to the ground state. 

We will start with the exact form of $\tilde \rho$, which is the state of one internal system in one copy. 
Let us note the following identity for two density matrices $\phi$ and $\psi$, 
\begin{equation}
\begin{split}
    \Tr_2(e^{i a S} & \phi \otimes \psi e^{-i a S}) = 
    \\
    & \cos^2(a) \phi+ \sin^2(a) \psi + i \cos(a)\sin(a)(\psi \phi-\phi \psi).
\end{split}
\label{TraceSWAPTrickExact}
\end{equation}
From the above, the action of $U_{2i,2i-1}$, as defined in Eq.~\eqref{Uij}, on two copies is given by setting $a=\pi/4$, and replacing $\phi\to e^{-i\e H}\phi e^{i\e H}$ and $\psi \to e^{i\e H}\phi e^{-i\e H}$, which gives 
\begin{equation}
    \begin{split}
    \Tr_2(& U  \phi \otimes \phi U^\dagger) = 
    \\
    & \frac{1}{2} \left(e^{-i\e H}\phi e^{i\e H}+  e^{i\e H}\phi e^{-i\e H}\right) +
    \\
    & \ \ \ \frac{i}{2} (e^{i \e H} \phi e^{-i 2 \e H}\phi e^{i \e H}-e^{-i \e H} \phi e^{i 2 \e H}\phi e^{-i \e H}).
\end{split}
\end{equation}
Let us now take $H=H_{int}\otimes 1$ and let $\phi$ be the maximally entangled state $\phi=\frac{1}{D}\sum_{i,j}|ii\rangle\langle jj|$, where $D$ is the Hilbert-space dimension of the internal systems. Tracing over the second internal system, we obtain
\begin{equation}
\begin{split}
    \label{rhoTilInter}
    \tilde \rho &= \frac{1}{D}+
    \\ 
    & \frac{i}{2D^2}\left(\Tr( e^{-i 2\e H_{int}})e^{i 2 \e H_{int}}   -\Tr( e^{i 2 \e  H_{int}})e^{-i 2 \e H_{int}}\right).
\end{split}
\end{equation}

The density matrix above is diagonal in the energy basis, and can be written as $\tilde \rho = \sum_i p_i \ket{E_i}\bra{E_i}$.
Using the swap trick provides access to expectation values under $\tilde \rho^n/\Tr(\tilde \rho^n)$. Notice also that the probabilities $p_i$ are functions of the energies, and two energy eigenstates having the same energy will have equal probabilities. Suppose $p_*$ is the largest probability, \ie $p_*\geq p_i$ for all $i$, and there is no degeneracy. We denote the second largest probability as $p_{**}=p_*e^{-\e \Delta}$. The probability of measuring $\ket{E_*}\bra{E_*}$, under $\tilde \rho^n/\Tr(\tilde \rho^n)$ is
\begin{equation}
    \begin{split}
        \Tr(\ket{E_*}\bra{E_*} \tilde \rho^n/\Tr(\tilde \rho^n)) & = \frac{p_*^n}{p_*^n+\sum_{i\neq*}p_i^n}
        \\
        & \geq \frac{p_*^n}{p_*^n+Dp_{**}^n}
        \\
        & = \frac{1}{1+D(p_{**}/p_*)^n}
        \\
        & \sim 1-D e^{-n \e \Delta}.
    \end{split}
\end{equation}

\begin{figure}
    \centering
    \includegraphics[width=0.45\textwidth]{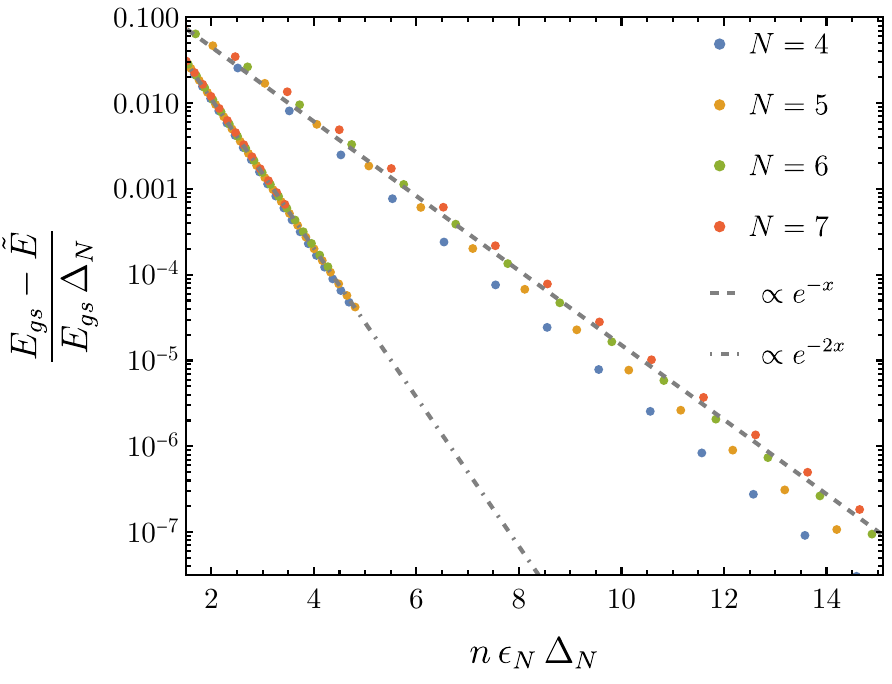}
    \caption{Numerical results for the single-layer protocol.
    $E_{gs}-\tilde E$ is the difference between the ground state energy and $\Tr(H_{int} \tilde \rho^n)/\Tr( \tilde \rho^n)$, where $H_{int} = -\sum_{i}(\sigma_z^i \sigma_z^{i+1}+\sigma_x^{i}+\sigma_z^{i})$, is a spin chain Hamiltonian of $N$ spins with periodic boundary conditions. $\Delta_N$ is the energy gap between the ground state and the first excited state. $\e_N$ is taken to be $0.19/N$, for one data set, and $0.03/N$ for the other. The results show that $\Tr(H_{int} \tilde \rho^n)/\Tr( \tilde \rho^n)\approx E_{gs}(1+\Delta_N e^{-a 
    n })$. When $\e$ is small enough, $a = 2 \e \Delta_N$, as $\tilde \rho \approx e^{-2 \e H_{int}}$ approximates the thermal state. Increasing $\e$ breaks the thermal state approximation, but can yield a faster apparent decay (larger fitted $a$).}
    \label{fig:swapTrick}
\end{figure}

In case there are $x$ eigenstates with probability $p_*$, similar arguments lead to
\begin{equation}
    \begin{split}
        \Tr(\ket{E_*}\bra{E_*} \tilde \rho^n/\Tr(\tilde \rho^n)) & = 
        \\
        & \geq \frac{1}{x}\frac{1}{1+\frac{D-x}{x}(p_{**}/p_*)^n}
        \\
        & \sim \frac{1}{x}-\frac{D-x}{x^2} e^{-n \e \Delta},
    \end{split}
\end{equation}
and therefore the state converges exponentially with $n$ to a uniform mixture of equal energy eigenstates, as demonstrated in Fig. \ref{fig:swapTrick}. 
This means that the energy expectation value measured in the large $n$ limit will be equal to an eigenvalue of the Hamiltonian, and different values of $\e$ can result in different eigenvalues, providing a method for spectroscopy.

Specifically, as we have already established, if $\e$ is small enough, meaning that $\e \|H_{int}\|\ll 1$, then $\tilde \rho \approx e^{-2 \e H_{int}}/\Tr(e^{-2 \e H_{int}})$. In that case, $\tilde \rho^n/\Tr(\tilde \rho^n)$ will converge to the ground state with $\Delta$ equals to twice the energy gap.

To analyze how the number of copies scales with the internal system's size, let us assume that $H_{int}$ is a gapped local Hamiltonian of $N$ qudits. In this case, $\|H_{int}\|$ and $\log(D)$ typically scale like $N$, and $\Delta$ is approximately constant. Therefore, for $D e^{-n \e \Delta} \ll 1$, $n$ needs to grow like,
\begin{equation}
    n \sim N^2,
\end{equation}
which is polynomial in the number of qubits.

Although the protocol appears efficient, its cost is shifted from the number of copies and gates to the number of measurements required to accurately evaluate $\Tr(H_{int} \tilde \rho^n)/\Tr( \tilde \rho^n)$ due to the exponential-in-$n$ small denominator. Nevertheless, it can be useful in cases where performing measurements is easier than increasing the number of copies or gates. Finally, the use of the swap trick for observables under $\Tr(H_{int} \tilde \rho^n)/\Tr( \tilde \rho^n)$ can be combined with the circuits in the previous sections to first prepare a small number of copies in the approximate thermal state with $\beta>2\e$. In this case, $\tilde \rho$ will be less mixed. Together with the smaller number of copies needed, $\Tr( \tilde \rho^n)$ will be closer to $1$ and fewer measurements will suffice.

\section{Outlook on experimental implementations}\label{sec:outlook}

Instead of relying only on digital quantum computers with universal control to execute our protocols, we will now discuss the requirements for implementations based on (hybrid-)analog quantum simulators with limited programmability.
Such settings constitute promising near-term applications that harness the high-fidelity implementation of quantum many-body models through large-scale analog quantum simulators in existing experimental setups. 
Our protocol can be naturally integrated with existing approaches, for example, after adiabatic state preparation, in order to further increase ground state fidelities.

Our protocols clearly require control over multiple copies of ideally identical systems. When based on Eq.~\eqref{newV}, however, they only require two operations:
\begin{enumerate}
    \item[($i$)] forward real-time evolution of individual copies, \ie an implementation of $e^{-i Ht}$ generated by a target Hamiltonian $H$ where the time can be chosen independently for each copy;
    \item[($ii$)] SWAP-rotations on pairs of copies, more specifically $e^{\pm i \frac{\pi}{4}S}$ where the operator $S$ swaps two individual copies.
\end{enumerate}
In fact, the second requirement can be replaced by the potentially simpler operation
\begin{enumerate}
    \item [($ii'$)] discrete controlled-SWAP operation on pairs of copies, \ie a conditional $CS = \ket{0}\bra{0} \otimes \mathbf{1} + \ket{1}\bra{1} \otimes S$ controlled by the state of an ancilla.
\end{enumerate}
This follows from the general identity
\begin{equation*}
\begin{quantikz}
\ket{0}         & \gate{H}& \ctrl{1} & \gate{X_\varphi} & \ctrl{1} & \gate{H} & \ket{0}\\
\ket{\psi_1} & & \gate[2]{S} &  & \gate[2]S & \rstick[2]{$e^{i\varphi S } \ket{\psi_1} \ket{\psi_2}$}\\
\ket{\psi_2} & & &  & &
\end{quantikz} 
\end{equation*}
where $H$ is the Hadamard gate and $X_\varphi = e^{i\varphi X} = \cos(\varphi) + i X \sin(\varphi)$ denotes an $X$-rotation.

When these requirements are met, both the tree and hedge~\footnote{The hedge circuit requires one small adjustment: within each layer, any copy not acted on by a $V$ gate is evolved for an additional $\e$ real time. This padding of the circuit equates the real time of the copies on which the $V$ gates act, such that also the acquired phases agree and the copies are symmetric w.r.t. SWAP (at leading order).} circuits (Fig.~\ref{fig:CircuitsCombined}) can be implemented using the $V$ gates defined in Eq.~\eqref{newV}.
Let us now make this more concrete for contemporary quantum simulation platforms and outline how to meet the two requirements.

\subsection{Atoms in optical lattices}
We consider neutral atoms trapped in optical lattices~\cite{Gross2017}.
Multiple copies can be realized~\cite{islam2015measuring,kaufman2016quantum}, e.g., as neighboring 1D sub-lattices of a global 2D lattice, or even as 2D layers of a global 3D lattice.

$(i)$: Atoms in optical lattices natively realize Fermi or Bose Hubbard models~\cite{Gross2017}.
Individual copies can be isolated by super-lattice potentials~\cite{bloch2012quantum}, such that tunneling between the sub-lattices is strongly suppressed.
By reshaping the potentials within each copy, together with the capability to turn off density-density interactions via Feshbach resonances~\cite{chin2010feshbach}, the native Hamiltonian dynamics can also be completely turned off, thereby enabling independent real-time evolution of individual copies.

$(ii)$: Previous theoretical works~\cite{pichler2016measurement,QVC2019} have also discussed implementations of the (controlled)-SWAP for such cold-atom setups.
In this context, $S$ is (up to potential phase factors) given by a local tunneling operation in the optical lattice.
The $CS$ gate can be realized, e.g., by combined photon-assisted tunneling with a Rydberg blockade mechanism that prevents tunneling iff an ancilla atom is in a Rydberg state.
Alternatively, sufficiently strong density-density interaction could, in principle, be employed to realize a controlled operation.
In the fermionic case, however, we emphasize that some care has to be taken due to non-trivial exchange statistics (see, e.g., the discussion in Ref.~\cite{pichler2013thermal}).
Nevertheless, our protocol only requires an intermediate level of control when comparing existing (analog) Fermi-Hubbard experiments to recently proposed (fully digital) fermionic quantum processors~\cite{gonzalez2023fermionic}. We also note the recent proposal of Ref.~\cite{gkritsis2025simulating}, where quantum-number-preserving variational ans\"atze for quantum chemistry are compiled into Hubbard-type operations in fermionic optical superlattices. In the context of low-energy state preparation, Ref.~\cite{gkritsis2025simulating} provides a complementary route that relies on variational optimization over a hardware-native ansatz and repeated energy measurements, rather than multi-copy manipulations. In chemistry applications, such variationally prepared low-energy states could also serve as improved initial states for the protocol developed here.

\subsection{Rydberg atom arrays}
As a second example, consider neutral atoms trapped in arrays of optical tweezers~\cite{browaeys2020many}.
As tweezers can be reconfigured essentially arbitrarily, this also enables control over independent copies in real space~\cite{bluvstein2022quantum}.

$(i)$: When excited to high-lying Rydberg states, atom arrays naturally implement a variety of spin models~\cite{browaeys2020many}, notably Ising-type and XY models with long-range van der Waals or dipolar interactions, respectively.
In such a setting, the single-spin operations are fully controlled by laser fields, while Rydberg interactions can also be turned on and off by transferring population from the Rydberg state to a hyperfine ground state~\cite{bluvstein2022quantum}.
In this way, individual copies can be independently time-evolved as required.

$(ii)$: To realize the required $S$ or $CS$ gates, one can rely on the demonstrated universal control over Rydberg atom arrays~\cite{bluvstein2022quantum}.
After the state of a copy is temporarily transferred into a hyperfine manifold, the $CS$ gate can be further decomposed into native one- and two-qubit gates.
Our protocols can then be executed in a hybrid setting where the time evolution is implemented in an analog mode, interleaved with digital SWAP operations.

\subsection{Other platforms}
We emphasize that our protocols can also be implemented on any other platform when requirements (i) and (ii), stated at the beginning of Sec.~\ref{sec:outlook}, are met.
For example, both superconducting qubits quantum computers, such as those in Ref.~\cite{google2025quantum}, and trapped-ion devices~\cite{blatt2012quantum,ransford2025helios} have demonstrated universal control which enables $(ii)$ through a digital synthesis of SWAP operations on isolated sub-systems as independent copies (see also Ref.~\cite{DBAC2025} for an implementation of a closely related protocol).
In fact, both platforms also enable $(i)$ through native analog Hamiltonian evolution as demonstrated in recent hybrid digital-analog experiments~\cite{andersen2025thermalization,foss2025progress}.
We conclude that our proposals are within reach of existing hardware.

\section{Summary}
In this work, we developed deterministic multi-copy circuits that approximate normalized imaginary-time evolution and can therefore be employed to prepare quantum many-body ground states. The construction of these circuits relies only on ingredients natural in several near-term architectures: the ability to store multiple copies of the system, implement real-time evolution under the target Hamiltonian on each copy, and carry out ancilla-controlled SWAP operations between copies. A central observation is that the dynamical algebra generated by these primitives gives access to the Hermitian interaction \( i\,(H\otimes 1 - 1\otimes H)\,S \), enabling unitaries of the form \(e^{\epsilon (H\otimes 1 - 1\otimes H)\,S}\). When applied to two copies, these unitaries reproduce, to leading order in \(\epsilon\), the effect of imaginary-time evolution, and additional copies allow larger effective imaginary times to be built up from such elementary steps.

We analyzed two circuit families whose width is set by the number of copies. The tree architecture admits a provable polynomial-in-depth convergence: for circuit depth \(n\), the error decreases as \(1/n\), but this comes at the cost of a width that grows exponentially with depth. More specifically, the worst-case upper bound in Eq.~\eqref{boundExp} scales exponentially in \(\beta \|H\|\), and hence exponentially with system size. Our numerical results, however, shown in Fig.~\ref{fig:three_panels}, suggest that for physically relevant systems, the scaling with system size can be substantially more favorable than the worst-case scenario. 

The hedge architecture, as supported by simulations, offers a promising heuristic alternative, achieving comparable accuracy with only polynomial width. We then find numerical evidence that modest mid-circuit post-selection can further accelerate convergence with practical success probabilities. We discussed how, for ground state preparation, circuit volume and the shot complexity of end-of-circuit observable (such as energy) estimation can be interchanged, and provided an explicit example demonstrating the tradeoff. Finally, we outlined an implementation route for platforms where multi-copy registers and SWAP-mediated couplings are available. These hybrid analog-digital protocols can already be implemented with existing quantum simulation platforms, both as an extra layer alongside other state-preparation methods when additional precision is needed, and in order to study thermal behavior.

From a theoretical perspective, a required direction would be to further explore the convergence of the protocol. Specifically, to study under which conditions the protocol with the hedge circuit indeed converges polynomially in the number of copies, and to further study what Hamiltonians allow for the number of copies to be polynomial in the system size.
It would be of importance to derive schemes that improve the convergence rate, for example, by tuning the gate parameter $\epsilon$ along the circuit. Given that the error of each gate from the exact imaginary time evolution depends on the state's energy spread (as can be seen by the variance term in Eq. \eqref{traceNormDisV5}), it is reasonable to expect that a varying $\e$ that is larger at later steps touching the first copy could speed up the protocol. Another interesting option would be to study further how limiting post-selection to specific locations along the circuit can improve the post-selection versions of the protocols.

The heuristic guideline used in the protocol was to treat the gates as if they do imaginary time evolution on two copies, one backward and one forward. However, the full ideal circuit in which we replace the gates in the circuits with the exact imaginary time evolution does not respect the exact symmetries of the actual circuit: given that the gates
are generated by swaps and single copy Hamiltonians, the actual circuit conserves any functional of the energy that is symmetric under all permutations of the copies, for example, the total energy. The symmetry argument above allows for the interrogation of the state of the rest of the copies - their energy and entanglement distributions -  to both optimize and understand the protocols' analytic behavior.
Finally, understanding the impact of noise on the protocol's efficacy will be important for realistic near-term implementations.

\acknowledgments

We would like to thank Reuben Wang, Henning Schloemer, Christian Kokail, and Robert Ott for helpful discussions.  
This work is supported by the European Union’s Horizon Europe research and innovation program under Grant Agreement No.~101113690 (PASQuanS2.1), the ERC Starting grant QARA (Grant No.~101041435), the Austrian Science Fund (FWF) (Grant No. DOI 10.55776/COE1), and the ERC Starting grant QS-Gauge (Grant No.~101220401).
Views and opinions expressed are however those of the author(s) only and do not necessarily reflect those of the European Union or the European Research Council Executive Agency.
Neither the European Union nor the granting authority can be held responsible for them. TS and HRS acknowledge support for this work from ITAMP, funded by the US National Science Foundation. TS acknowledges support from the VATAT
Outstanding Postdoctoral Fellowship in Quantum Science
and Technology. TS is grateful for the generous support from the RH Growth Foundation.

\newpage
\appendix

\onecolumngrid

\section{Useful SWAP identities and the core equations}
\label{SwapIdentitiesAppendix}

For completeness, we collect here the SWAP identities used in
Eqs. \eqref{SwapEvoForward} \eqref{SwapEvoBackward}, \eqref{Utwo}, and \eqref{TraceSWAPTrickExact}. Let $S$ be the SWAP operator on two
identical Hilbert spaces. For arbitrary operators $A$ and $B$,
\begin{equation}
    S^\dagger=S,
    \qquad S^2=\mathbf{1},
    \qquad
    S(A\otimes B)S=B\otimes A,
\end{equation}
and
\begin{equation}
    \Tr_2[(A\otimes B)S]=AB,
    \qquad
    \Tr_2[S(A\otimes B)]=BA, \qquad \Tr_1[(A\otimes B)S]=BA, \qquad \Tr_1[S(A\otimes B)]=AB.
\label{PartialTraceIdentities}
\end{equation}
Since $S^2=\mathbf{1}$, a series expansion shows that the finite SWAP rotation can be written as
\begin{equation} \label{SWAProtation}
    e^{iaS}=\cos(a)\mathbf{1}+i\sin(a)S.
\end{equation}
If $\Tr(A) = \Tr(B)=1$, as in the case $A$ and $B$ are density matrices, the above two equations give,
\begin{align}
    & \Tr_2\left(
    e^{iaS}A\otimes B e^{-iaS}
    \right)
    =
    \cos^2(a)A +\sin^2(a)B
    +
    i\cos(a)\sin(a)(BA-AB)
    \\
    & \Tr_1\left(
e^{iaS}A\otimes B e^{-iaS}
\right)
=
\cos^2(a)B
+
\sin^2(a)A
+
i\cos(a)\sin(a)(AB-BA).
    \label{ExactPartialSwap}
\end{align}
This is the identity used to derive Eq. \eqref{TraceSWAPTrickExact}.

Setting $a=\sqrt{\e}$, $A=\rho_1$ and $B=\rho_2$, a small $\e$ expansion reduces to 
\begin{align}
 \label{Eq4Again}   \Tr_2\left(e^{i\sqrt{\e}S}\rho_1\otimes\rho_2
    e^{-i\sqrt{\e}S}
    \right)
    = &
    \rho_1+i\sqrt{\e}[\rho_2,\rho_1]
    +
    \e(\rho_2-\rho_1)
    +O(\e^{3/2}),
    \\
    \Tr_1\left(e^{i\sqrt{\e}S}\rho_1\otimes\rho_2
    e^{-i\sqrt{\e}S}
    \right)
    = &
    \rho_2+i\sqrt{\e}[\rho_1,\rho_2]
    +
    \e(\rho_1-\rho_2)
    +O(\e^{3/2})
    ,
    \label{Eq4AgainRev} 
\end{align}
which reproduces Eq.~\eqref{SwapEvo}. 
For
\begin{equation}
    \rho_2=e^{-i\sqrt{\e}H}\rho_1e^{i\sqrt{\e}H}
    =
    \rho_1-i\sqrt{\e}[H,\rho_1]+O(\e),
    \label{rho2asrho1}
\end{equation}
the $\e (\rho_2-\rho_1)$ term of Eq. \eqref{Eq4Again} contributes only at order $\e^{3/2}$, and therefore we obtain Eq. \eqref{SwapEvoForward}.
Similarly, if Eq. \eqref{rho2asrho1} holds, then $\rho_1 = e^{i\sqrt{\e}H}\rho_2 e^{-i\sqrt{\e}H}$, and Eq. \eqref{Eq4AgainRev} gives Eq. \eqref{SwapEvoBackward}.

We finally explain the second equality in Eq. \eqref{Utwo}. With
\begin{equation}
    K=H\otimes\mathbf{1}-\mathbf{1}\otimes H,
\end{equation}
the SWAP operator satisfies
\begin{equation}
    SKS=-K,
    \qquad
    SK=-KS.
\end{equation}
Using Eq. \eqref{SWAProtation}, it follows that
\begin{equation}
    e^{i\frac{\pi}{4}S}
    K
    e^{-i\frac{\pi}{4}S}
    =
    -iKS.
\end{equation}
Consequently,
\begin{align}
    e^{i\frac{\pi}{4}S}
    e^{-i\e K}
    e^{-i\frac{\pi}{4}S}
    &=
    \exp\left[
    -i\e
    e^{i\frac{\pi}{4}S}
    K
    e^{-i\frac{\pi}{4}S}
    \right]
    \nonumber\\
    &=
    e^{-\e KS}.
\end{align}
Although the last exponent contains no explicit factor of $i$, the
operator is unitary because
\begin{equation}
    (KS)^\dagger=SK=-KS.
\end{equation}

\section{Errors of the elementary gates}
\label{ElementaryGateErrors}

Here, we give the expansions leading to
Eqs.~\eqref{traceNormDis}, \eqref{traceNormDisV5}, and
\eqref{traceNormDisVtilde}. We define
\begin{equation}
\Phi_z=\phi_z\otimes\phi_z,
\qquad
\langle X\rangle_z=\Tr(X\Phi_z),
\qquad
K=H\otimes\mathbf{1}-\mathbf{1}\otimes H.
\end{equation}
For a pure state $\ket{\Phi_z}$, the density matrix $\Phi_z = \ket{\Phi_z}\bra{\Phi_z}$, and a  Hermitian operator $A$, the trace distance satisfies,
\begin{equation}
\frac{1}{2}\left\|i[A,\Phi_z]\right\|_1 =
\frac{1}{2}
\left\|
[A-\langle A\rangle_z,\Phi_z]
\right\|_1
=
\frac{1}{2}
\left\|
\{A-\langle A\rangle_z,\Phi_z\}
\right\|_1 = 
\sqrt{
\langle A^2\rangle_z-\langle A\rangle_z^2
}.
\label{PureStateNormIdentities}
\end{equation}
This can be seen, for example, by noticing that $(A - \langle A\rangle_z)\ket{\Phi_z}$ is orthogonal to $\ket{\Phi_z}$, and therefore $\sqrt{
\langle A^2\rangle_z-\langle A\rangle_z^2
}^{-1}i[A-\langle A\rangle_z,\Phi_z]$, and $\sqrt{
\langle A^2\rangle_z-\langle A\rangle_z^2
}^{-1}\{A-\langle A\rangle_z,\Phi_z\}$ in matrix form are the Pauli $y$ and Pauli $x$ matrices, respectively, which have trace norm that equals $2$. 

For the gate $W$, expanding the two states appearing in
Eq.~\eqref{traceNormDis}, gives
\begin{equation}
\begin{aligned}
&\phi_{z+\e}\otimes\phi_{z-\e+i\sqrt{\e}}
-W\Phi_z W^\dagger
=
i\e^{3/2}
[K+\frac{1}{2}K^2,\Phi_z]
+O(\e^2).
\end{aligned}
\label{WLeadingDifference}
\end{equation}
It follows from Eq. \eqref{PureStateNormIdentities} that
\begin{equation}
\begin{aligned}
&\frac{1}{2}
\left\|
\phi_{z+\e}\otimes\phi_{z-\e+i\sqrt{\e}}
-W\Phi_z W^\dagger
\right\|_1
 =
D_z(H)\e^{3/2}+O(\e^2),
\end{aligned}
\end{equation}
where
\begin{equation}
\begin{aligned}
D_z(H)^2
&=
\left\langle
\left(K+\frac{1}{2}K^2\right)^2
\right\rangle_z
-
\left\langle
K+\frac{1}{2}K^2
\right\rangle_z^2.
\end{aligned}
\label{DzDefinition}
\end{equation}
In particular, $D_z(H)=0$ when $\phi_z$ is an
eigenstate of $H$.

For the gate $U$, the corresponding expansion is
\begin{equation}
\begin{aligned}
&\phi_{z-\e}\otimes\phi_{z+\e}
-U\Phi_zU^\dagger
=
\e^2
\left\{
K^2-\langle K^2\rangle_z,\Phi_z
\right\}
+O(\e^3),
\end{aligned}
\label{ULeadingDifference}
\end{equation}
and from Eq. \eqref{PureStateNormIdentities}, we obtain 
\begin{equation}\label{derOfUerror}
\begin{aligned}
&\frac{1}{2}
\left\|
\phi_{z-\e}\otimes\phi_{z+\e}
-U\Phi_zU^\dagger
\right\|_1
 =
\e^2
\sqrt{
\langle K^4\rangle_z
-
\langle K^2\rangle_z^2
}
+O(\e^3)
=
\sqrt{\text{Var}_z(K^2)}\e^2+O(\e^3),
\end{aligned}
\end{equation}
which gives Eq.~\eqref{traceNormDisV5}.

For the gate $V$, defined in Eq. \eqref{newV}, the unitary invariance of the trace norm gives that 
\begin{equation}
    \frac{1}{2}
\left\|
\phi_{z-\e+i\e}\otimes\phi_{z+\e+i\e}
-V\Phi_zV^\dagger
\right\|_1 = \frac{1}{2}
\left\|
\phi_{z-\e}\otimes\phi_{z+\e}
-e^{i\e(H \otimes 1+1 \otimes H) }V\Phi_zV^\dagger e^{-i\e(H \otimes 1+1 \otimes H)  }
\right\|_1.
\end{equation}

As $H \otimes 1+1 \otimes H$ commutes with the SWAP, $e^{i\e(H \otimes 1+1 \otimes H) }V = U$, and
\begin{equation}
    \frac{1}{2}
\left\|
\phi_{z-\e+i\e}\otimes\phi_{z+\e+i\e}
-V\Phi_zV^\dagger
\right\|_1 = \frac{1}{2}
\left\|
\phi_{z-\e+i\e}\otimes\phi_{z+\e+i\e}
-U \Phi_z U^\dagger
\right\|_1.
\end{equation}
Therefore, with Eq. \eqref{derOfUerror}, we derive Eq. \eqref{traceNormDisVtilde}.

\section{Tree circuit error bounds}\label{TreeErrorBound}

In the following, we shall derive Eq.~\eqref{boundExp}, an upper bound on the trace distance between the exact imaginary time evolved state, and the tree circuit's output $\tilde \phi_1^{(n)}$. The evolution along the layers of the circuit of the first copy's state follows the relation in Eq.~\eqref{TreeRecurRel}. Let us define $\phi_1^{(l)}=\phi_{l\e}$ as the exact imaginary time evolved state on the first copy, with imaginary time $l  \e$, \ie at each layer, $\phi_1^{(l)}$, is imaginary time evolved with imaginary time $\e$. Let us also define $\delta_1^{(l)}=\tilde \phi_1^{(l)}-\phi_1^{(l)}$, as the difference between the states. Twice the trace distance at layer $l+1$, follows the upper bound below: 
\begin{equation}
    \begin{split}
        \| \delta_1^{(l+1)}\|_1 & \equiv \|  \phi_1^{(l+1)}  -\tilde \phi_1^{(l+1)}\|_1 
        \\
        & = \| \phi_1^{(l+1)} -\Tr_2(U \, \tilde \phi_1^{(l)}\otimes \tilde \phi_1^{(l)} U^\dagger)\|_1 
        \\
        & = \| \phi_1^{(l+1)} -\Tr_2(U \,  (\phi_1^{(l)}+\delta_1^{(l)})\otimes (\phi_1^{(l)}+\delta_1^{(l)}) U^\dagger)\|_1 
        \\
        & \leq \| \phi_1^{(l+1)} -\Tr_2(U \,  \phi_1^{(l)}\otimes  \phi_1^{(l)} U^\dagger)\|_1 +\| \Tr_2(U (  \delta_1^{(l)}\otimes (\phi_1^{(l)}+\delta_1^{(l)})+  \phi_1^{(l)}\otimes \delta_1^{(l)}) U^\dagger)\|_1 
        \\
        & \leq 2\sigma_l(K^2) \e^2 +\| \Tr_2(U (  \delta_1^{(l)}\otimes (\phi_1^{(l)}+\delta_1^{(l)}))U^\dagger)\|_1+\|\Tr_2(  U\phi_1^{(l)}\otimes \delta_1^{(l)} U^\dagger)\|_1
        \\
        & \leq 2\sigma_l(K^2)  \e^2 +\|  \delta_1^{(l)}\|_1 \, \|(\phi_1^{(l)}+\delta_1^{(l)})\|_1+\|\Tr_2(  U\phi_1^{(l)}\otimes \delta_1^{(l)} U^\dagger)\|_1
        \\
        & \leq 2\sigma_l(K^2)  \e^2 +\|  \delta_1^{(l)}\|_1 +\|\Tr_2(  U\phi_1^{(l)}\otimes \delta_1^{(l)} U^\dagger)\|_1,
    \end{split}
\end{equation}
where we have used Eq.~\eqref{TreeRecurRel} in the second line (suppressing the subscripts denoting on which copies $U$ acts), the triangle inequality in the third line, and Eq.~\eqref{traceNormDisV5} in the fifth line. In the sixth line, we have used the fact that the trace norm is monotonic under a partial trace, \ie $\|\Tr_2(*)\|_1\leq \|*\|_1$, that it is invariant under unitary transformations, and that $\| A \otimes B\|_1 = \|A\|_1 \, \|B\|_1$. Finally, in the seventh line, we have used the fact that $\phi_1^{(l)}+\delta_1^{(l)} = \tilde \phi_1^{(l)}$ is a normalized state, and therefore has a unit trace norm.

Let us now use the $\e$ expansion of the unitary, $U=1+\e K S - \e^2/2 \, K^2 + O(\e^3)$,\footnote{ We note that the sign of $\e$ here is opposite to the definition of $U$ in Eq. \eqref{Utwo}. This is because we (arbitrarily) chose to evolve the first copy from $\phi_0$ to $\phi_\beta$, and not to $\phi_{-\beta}$ in the error bound derivation.   }
\begin{equation}
    \begin{split}
         \| & \delta_1^{(l+1)}\|_1 
          \\
          & \leq 2\sigma_l(K^2)  \e^2 +\|  \delta_1^{(l)}\|_1 +\|\Tr_2( \phi_1^{(l)}\otimes \delta_1^{(l)} +\e [K S,\phi_1^{(l)}\otimes \delta_1^{(l)}] + \e^2 K S \phi_1^{(l)}\otimes \delta_1^{(l)} S K  -\frac{\e^2}{2} \{K^2,\phi_1^{(l)}\otimes \delta_1^{(l)}\})\|_1+O(\e^3)
        \\
        & \leq 2\sigma_l(K^2)  \e^2 +\|  \delta_1^{(l)}\|_1 +\|\Tr_2(\e [K S,\phi_1^{(l)}\otimes \delta_1^{(l)}] + \e^2 K S \phi_1^{(l)}\otimes \delta_1^{(l)} S K  -\frac{\e^2}{2} \{K^2,\phi_1^{(l)}\otimes \delta_1^{(l)}\})\|_1+O(\e^3)
        \\
        & \leq 2\sigma_l(K^2)  \e^2 +\|  \delta_1^{(l)}\|_1 +(2 \e  \|
        KS\|+\e^2 \|KS\| \|SK\|+\e^2 \|K^2\|)\|\phi_1^{(l)}\otimes \delta_1^{(l)}\|_1  + O(\e^3)
        \\
        & \leq 2\sigma_l(K^2)  \e^2 +\|  \delta_1^{(l)}\|_1 +(2 \e  \|
        K\|+\e^2 B_K)\|\delta_1^{(l)}\|_1  + O(\e^3)
        \\
        & =  2\sigma_l(K^2)  \e^2  +(1+2 \e  \|
        K\|+\e^2 B_K)\|\delta_1^{(l)}\|_1  + O(\e^3)
        \\
        & =  2\sigma_{l_*}(K^2)  \e^2  +(1+4 \e  \|
        H\|+\e^2 B_K)\|\delta_1^{(l)}\|_1  + O(\e^3),
    \end{split}
\end{equation}
where we have used in the third line the fact that $\Tr(\delta^{(l)}_1)=0$. To obtain the fourth line, we again used the triangle inequality together with the monotonicity of the trace norm under the partial trace, and also that the trace norm satisfies H{\" o}lders inequality, $\|AB\|_1 \leq \|A\| \, \|B\|_1$, where $\|*\|$ is the operator norm. In the fifth line, we note again that the trace norm factorizes under tensor products, and that states have unit norms; in addition we have used thw invariance of the operator norm under unitary operators like $S$ and define $B_k = \|K\|^2+\|K^2\|$, $\sigma_{\t_*}(K^2)=\max_l \sigma_{\t}(K^2)$. Finally, we have used the triangle inequality to obtain the last line.

By solving the recurrence relation above and setting $\e = \beta/n$, where $n$ is the total number of layers acting on the first copy, we obtain the result in Eq.~\eqref{boundExp}.

\end{document}